\documentclass[aps,showpacs,prl,twocolumn,superscriptaddress]{revtex4}
\usepackage{graphicx}
\usepackage{epsfig}
\newcommand{\beq}{\begin{equation}}
\newcommand{\eeq}{\end{equation}}
\newcommand{\bea}{\begin{eqnarray}}
\newcommand{\eea}{\end{eqnarray}}
\newcommand{\pdag}{{\phantom{\dagger}}}

\begin{document}

 \bibliographystyle{apsrev}

\title{Current fluctuations of an interacting quantum dot }

\author{Markus Kindermann}
\affiliation{Department of Physics, Massachusetts Institute of Technology,
Cambridge MA 02139, USA }
\author{Bj{\"o}rn Trauzettel}
\affiliation{ Laboratoire de Physique des Solides, Universit\'e Paris-Sud,
91405 Orsay, France}

\date{\today}
 
\begin{abstract}
We calculate the counting statistics of electron transfer through an open quantum 
dot with charging interaction. A dot  that is connected to leads by two 
single-channel quantum point contacts in an in-plane magnetic field is described 
by a Luttinger liquid with impurity at the Toulouse point. We find that the fluctuations 
of the current through  this conductor exhibit distinctive interaction effects. Fluctuations 
saturate at high voltages, while the mean current increases linearly with 
the bias voltage. All cumulants higher than the second one reach at large bias a 
temperature independent limit.
\end{abstract}
\maketitle

The statistical distribution of current fluctuations in nanoscale conductors, 
the so-called ``counting statistics'', has received considerable attention over 
the past decade - theoretically \cite{Lev93,QN03} as well as experimentally \cite{Reu03}. 
The theory of this distribution is well developed for non-interacting conductors. Effects of   
electron-electron interactions, however, have only been studied in a few limiting cases so far. 
Interacting conductors in the tunneling limit of weak transmission have been shown to 
exhibit Poissonian current fluctuations \cite{Lev01}. Also for weakly interacting 
conductors at arbitrary transmission the statistics of charge transfer has been found to be 
qualitatively 
identical to that of a non-interacting system - it is multinomial \cite{Kin03,Bag03}. 
In a conductor that has arbitrary transmission combined with strong interactions one 
expects more profound changes of that statistics. This is indicated by a classical  
analysis  \cite{Kin02} that has already revealed qualitative changes of 
the distribution of current fluctuations in that regime. In this Letter, we confirm this expectation by  a 
nonperturbative quantum mechanical analysis of non-Gaussian current fluctuations in a 
strongly interacting nanostructure at arbitrary transmission.
  
An interacting nanostructure is at low energy scales described by the model of a 
non-interacting conductor in an electromagnetic environment, that is a quantum conductor 
with an effective series resistor \cite{NazIng}. Recently, 
Safi and Saleur \cite{Saf03} have shown, that for a single-channel conductor this 
model maps onto a Luttinger liquid (LL) with  impurity. The statistics of fluctuations 
for this model has been obtained at zero temperature  
in the context of tunneling in fractional quantum Hall samples \cite{Sal01}. 
From a perturbative analysis  one expects that 
the current $I$ in systems described by a LL scales with a power of the 
applied voltage $V$ \cite{Kan92}. 
In the limit of a strong impurity, one has $I \propto V^{2/g-1}$, 
where $g$ is the interaction parameter of the LL. The 
current $I_B$ that is backscattered by a weak barrier obeys $I_B\propto V^{2g-1}$. 
Evidently then, $g=1/2$ is a special point in parameter space: The backscattered 
current becomes independent of the applied voltage.  
By virtue of the mapping of Ref.\ \cite{Saf03}, this unusual behavior should 
be observable in a quantum conductor with series resistance $1/G_0$, where 
$G_0=e^2/h$ is the conductance quantum. An open quantum point contact (QPC) is a 
natural realization of such a series resistor.
\begin{figure}
\includegraphics[width=8cm]{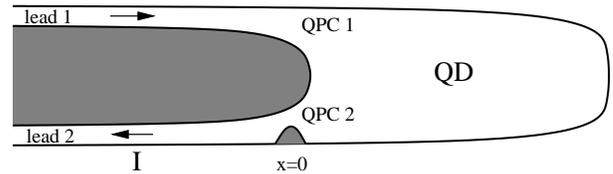}
\caption{QD connected to two leads by single-channel point contacts QPC $1$ 
and QPC $2$. QPC $2$ introduces backscattering. The current $I$ of charge transferred   
from lead $1$ to lead $2$ is measured.} 
\label{fig1}
\end{figure}

Motivated by this, we study transport through the interacting quantum dot (QD) shown in 
Fig.\ \ref{fig1}. It is connected to two leads via 
single-channel QPCs. One of them (QPC $1$) is open such as to act as the desired series 
resistor  to a scatterer in QPC $2$. An in-plane magnetic field allows only electrons in 
one particular spin state to  pass the QPCs. We address the incoherent case when inelastic 
relaxation inside the QD is faster than electron escape. The charging energy of the QD is assumed to be much larger than the temperature $T$ and the applied voltage. In our analysis, we employ 
techniques that have been developed in the
context of the Coulomb blockade \cite{Fle93,Mat95,Fur95,Ale02} and charge pumping in almost 
open QDs \cite{Ale98}. After bosonization, the QD in Fig.\ \ref{fig1} is at low energies, 
indeed, described by a LL at $g=1/2$ (the Toulouse point). This allows us to study its 
transport behavior nonperturbatively by refermionization.  We find as expected that $I_B$, 
the current backscattered by QPC $2$, saturates with increasing voltage.  It is, however, 
hidden in the total current $I=I_0-I_B$ under a large background  current 
$I_0 = g G_0 V  $ of charge carriers that are not backscattered. Analyzing also the 
fluctuations of the current through the structure, we find that the same physics leads 
to a saturation with increasing $V$ of non-Gaussian current 
fluctuations as well. No similar background as $I_0$ is present for  these
fluctuations, since all electrons that are not scattered produce only Gaussian 
thermal noise.  This, together with recent experimental advances in the detection of 
non-Gaussian current fluctuations \cite{Reu03}, brings the observation of this 
counterintuitive saturation effect within experimental reach. 
Remarkably, all cumulants higher than the second one are entirely 
{\em temperature independent} in the limit $V\to\infty$. This is in marked contrast with the behavior of non-interacting conductors. 

We assume the QD large enough for electrons to relax inelastically before they escape the dot.    In this case, the QD can be modeled by one  
one-dimensional electron mode for each QPC \cite{Fur95,Ale02} subject to a charging 
interaction inside the QD (at $x>0$, {\it cf.} Fig.\ \ref{fig1}). At the energies 
$\varepsilon$ of interest, that are smaller 
than the Zeeman energy of electrons in the applied magnetic field, only electrons of one 
particular spin  projection contribute to transport through the QPCs. 
 For typical conductors the electronic 
spectrum at the relevant energies can be linearized and the Hamiltonian is diagonalized 
by bosonization. The QD is then described by two bosonic fields  
$\theta_j$ and $\phi_j$ for every lead $j$ that describe electron density fluctuations 
\cite{Mat95,Gog}. Moreover, for energies below the charging 
energy $E_c$ of the QD, $\varepsilon \ll E_c$, the ``charge'' mode $\theta_1+\theta_2$ is 
pinned and the system is described by an effective Hamiltonian $H_{\rm eff}$  for the   
``transport'' modes $\theta_s=(\theta_1-\theta_2)/2$ and $\phi_s=(\phi_1-\phi_2)/2$.  
$\theta_s(0)$  is the operator of the charge $Q$ transported through the QD, 
$Q=e\theta_s(0)/\pi$. We have ($\hbar \equiv k_B \equiv 1$)
\bea
H_{\rm eff}&=& \int dx \, \left\{ \frac{v_F}{2\pi} \left[ 
\frac{1}{2} (\partial_x \phi_s)^2+  2(\partial_x \theta_s)^2 \right] \right. 
\nonumber \\
&& + \left. \delta(x)  \left( 2\lambda \cos 2\theta_s+ \frac{eV}{\pi} \theta_s\right)
\right\} .   
\eea
$\lambda$ is the backscattering strength of QPC $2$. 
 
To quantify the current $I$ through the QD and its fluctuations, we compute a 
generating function ${\cal Z}$ that generates moments of the charge 
$Q_{\tau}=\int_0^{\tau}{dt\,I(t)}$ transferred during time $\tau$,
\begin{equation} \label{funcdiff}
{\cal Z}(i \xi)=\sum_k \frac{\xi^k}{k!} \left\langle Q_{\tau}^k\right\rangle =
\exp\left[\tau \sum_k \frac{\xi^k}{k!} {\cal C}_k  \right] ,
\end{equation}
where $\tau {\cal C}_k$ is the $k$th cumulant of the distribution.
Writing ${\cal Z}$ as a Keldysh path integral \cite{Kam01} and integrating 
out $\phi_s$ as well as all  modes $\theta_s(x)$ at $x\neq 0$, we arrive at 
\bea \label{Zb}
{\cal Z}(\xi)&=& \int {\cal D} \theta\,  e^{{\cal S}_0 + {\cal S}_s-i 
\int{dt\,   2 \lambda (\cos 2\theta^+ - \cos 2\theta^-) } }
\eea
with the free action
\[
{\cal S}_0=\int{\frac{d\omega}{\pi^2}\, \omega 
\left[4 N \theta^{+ *} \theta^{-}-(2N+1) (|\theta^+|^2+|\theta^-|^2)   \right]}
\]
for fields $\theta^{\pm}$ corresponding to  the mode $\theta_s(0)$ on the 
forward ($\theta^+$) and the backward  ($\theta^-$) part of the Keldysh time  contour.  
$N=[\exp (\omega/T)-1]^{-1}$ is the Bose-Einstein distribution. 
The source term
\beq
{\cal S}_s = i \int_0^{\tau}{dt\, \left[\frac{e\xi}{2\pi} 
\partial_t(\theta^{+}+\theta^-) - \frac{eV}{\pi} (\theta^+-\theta^-)\right] }
\eeq
couples  $V$ and $\xi$ to charge $e\theta/\pi$ and current 
$-e\partial_t\theta/\pi$, respectively \cite{footnote}. 
We eliminate ${\cal S}_s$ by the change of integration variables 
$\theta^{\pm}(t)\to \theta^{\pm}(t) - e\tilde{V}t/4 \pm e\xi/8$, where  
$\tilde{V}= V - 2i\xi T $. The resulting ${\cal Z}$ can equivalently be obtained 
by integrating out all modes $\phi(x)$ and $\theta(x)$ at $x\neq 0$ in a path 
integral corresponding to the Hamiltonian expression 
\beq \label{Ham}
{\cal Z}(\xi)= {\cal Z}_0(\xi)\left\langle {\cal T}_{\pm} \,e^{{i}
\int_0^{\tau}{dt\,
  {H}^-(t)}} \,e^{-{i}
\int_0^{\tau}{dt\,
  {H}^+(t)}} \right\rangle
\eeq
with time-dependent Hamiltonians
\bea
 H^{\pm}(t)&=& \int dx \, \left\{ \frac{v_F}{2\pi} \left[ \frac{1}{2} 
 (\partial_x \phi)^2+ 2(\partial_x \theta)^2\right] \right. \nonumber \\
&&\left. +  2\lambda \delta(x)   \cos \left(2\theta\pm \frac{1}{4} e\xi - 
\frac{1}{2} e \tilde{V}t\right) \right\}.
\eea
${\cal T}_{\pm}$ orders operators along the Keldysh contour and 
\beq
{\cal Z}_0(\xi) = \exp\left[\tau \frac{e^2 }{4\pi}(-i V \xi - T \xi^2)\right]
\eeq
is the generating function in the absence of backscattering.
We now follow the standard procedure \cite{Gog} to refermionize 
$H^{\pm}$. For this we define new fields
\beq 
\phi_{\pm}(x) =  \frac{1}{2} [\phi(x)  \mp   \phi(-x)]+ \theta(x) \pm    \theta(-x)
\eeq
and a chiral Fermion $b=a^{-1/2} \exp (i\phi_+)$ with a short distance cutoff $a$. 
$H^{\pm}$ then have a representation as non-interacting Hamiltonians for $b$. 
Following Matveev \cite{Mat95} we introduce a Majorana 
fermion $d+d^{\dagger}$ and define new fermion operators $c$ by the relation $b=(d+d^{\dagger}) c $. This brings 
 $H^{\pm}$ into a quadratic form in fermion operators,
\bea \label{Hferm}
&& H^{\pm}(t)= \int{\frac{dk}{2\pi} \,\Bigl\{v_F k\, c_k^{\dag} c^{\pdag}_k  } \nonumber \\
&&\;\;\;\; \mbox{} + \sqrt{a} \lambda \left[ (d+d^{\dagger}) c_k 
e^{\pm ie\xi/4 -i e \tilde{V} t/2}   + h.c.\right] \Bigr\} .
\eea
Inserting Fermion coherent states, we rewrite  Eq.\ (\ref{Ham}) in refermionized 
form as a path integral \cite{Neg}. We again integrate out all modes 
$c(x\neq 0)$ and are left with an integral over vector fields 
$c=(c^+(0),c^-(0))$ and $d=(d^+,d^-)$,
\bea \label{Zpath}
&&{\cal Z}(\xi)={\cal Z}_0(\xi)\int{ {\cal D} c \, {\cal D} d \, 
\exp\Bigl\{ -i\int{\frac{d\omega}{2\pi}\,\Bigl[ d^{*} G_d^{-1} d }} \nonumber\\
&& \mbox{}+  c^{*} G_c^{-1} c  + \sqrt{a} \lambda 
\left(  (d+d^{*}) e^{ie\xi\tau^z/4} \tau^z c+ h.c. \right) \Bigr]\Bigr\}, \nonumber \\
\eea
where $\tau^z$ is the third Pauli matrix. 
$G_d(\omega) = -i \langle {\cal T}_{\pm} d(\omega) d^{\dag}(\omega)\rangle =  -\tau^z/\omega$ 
is the Green function of $d$  and $G_c$ that of $c$ corresponding to the 
Hamiltonian (\ref{Hferm}) at $\lambda=0$. The time dependence of the scattering term has 
been removed by a gauge transformation $c \to \exp(i e \tilde{V}t/2) \, c$ that shifts 
the frequency of $G_c$. Since $G_c$ is an electron Green function $G_c(x,x')$ 
evaluated at coinciding spatial coordinates  $x=x'=0$, it is linearly related to the 
semiclassical Keldysh Green function of $c$
\begin{equation}
 G_{s}=  \left(
\begin{array}{cc} 1 - 2 f & 2 f \\ 2(1 - f) & 2
f-1 \end{array} \right),
\end{equation} 
$G_c= G_s/4i v_F$. Here, $f(\omega)=[\exp(\omega-e\tilde{V}/2)/T+1]^{-1}$ is the Fermi 
distribution function after the gauge transformation. 

The action in Eq.\ (\ref{Zpath}) is diagonal in 
frequency and ${\cal Z}$ consequently factorizes into contributions from different 
frequencies.  The Gaussian integrals result in
\bea \label{stat}
&&\ln{\cal  Z}(\xi) = \ln{\cal Z}_0(\xi)+ \tau \int_0^{\infty}{ \frac{d\omega}{2\pi}\, 
\ln\Bigl\{1+\frac{T_B^2}{\omega^2+T_B^2} } \nonumber \\
&& \left[ (e^{ie\xi}-1)f^+(1-f^-)+ (e^{-ie\xi}-1)f^-(1-f^+)\right]\Bigr\}  \nonumber \\
\eea
with $f^+=f$, $f^-(\omega)=1-f(-\omega)$, and $T_B = a \lambda^2/2 v_F$.

We remark that one arrives at an equivalent result by applying the quasiparticle 
formalism developed by Fendley, Ludwig, and Saleur \cite{fendl95} to a LL at $g=1/2$.   

From Eq.\ (\ref{stat}) we find for the first three cumulants
\begin{eqnarray}
{\cal C}_1 &=&  {e^2\over 4\pi} V \left[1-{2 T_B\over e V} \; {\rm Im} \; 
\psi\left({1\over 2}+ 
{2 T_B + i e V\over 4\pi T}\right)\right], \label{c1exakt} \\
{\cal C}_2 &=& 2 T{d {\cal C}_1 \over dV} - \frac{e}{2} T_B 
\coth\left({eV\over 2 T}\right) {d {\cal C}_1 \over dT_B} \nonumber \\
&+& T  T_B{d^2 {\cal C}_1 \over dVdT_B}, \label{c2exakt} \\
{\cal C}_3 &=& \frac{1}{2} T T_B \frac{d^2 {\cal C}_2}{dV dT_B}
- \frac{e}{2} T_B \coth\left(\frac{eV}{2T}\right) \frac{d{\cal C}_2}{dT_B} \nonumber \\
&+& 2T \frac{d{\cal C}_2}{dV} + \frac{e}{2} T T_B \coth\left(\frac{eV}{2T} \right) 
\frac{d^2{\cal C}_1}{dVdT_B} \nonumber \\
&+& \frac{e^2}{4} T_B \sinh^{-2}\left(\frac{eV}{2T} \right) \frac{d{\cal C}_1}{dT_B} , 
\label{c3exakt}
\end{eqnarray}
where $\psi$ is the digamma function. At zero temperature, 
Eqs.~(\ref{c1exakt}) - (\ref{c3exakt}) obey the known relations between the higher order 
cumulants and ${\cal C}_1$ derived in Ref.\ \cite{Sal01}. 
 
\begin{figure}
\vspace{1.0cm}
\begin{center}
\epsfig{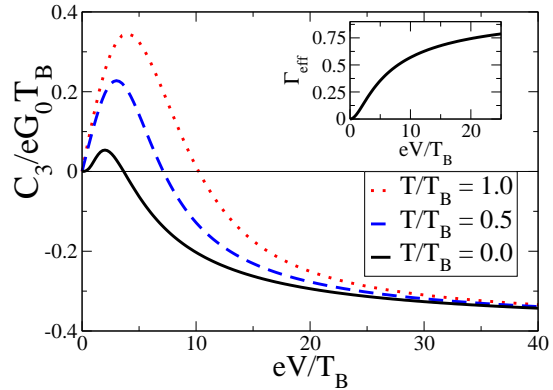}
\caption{\label{fig_largeV} ${\cal C}_3$ is plotted in units of $ e G_0 T_B$ versus
$eV/T_B$ for different temperatures.
For large voltages, ${\cal C}_3$ saturates at the temperature independent  
value ${\cal C}^{\infty}_3=  -(\pi e/8)G_0 T_B$. The inset shows that,
at zero temperature, ${\cal C}_3$ changes sign at an effective transmission 
$\Gamma_{\rm eff}\approx 0.26$.}
\end{center}
\end{figure}

We first analyze the limit of  a large voltage $eV\gg T,T_B$ (while $eV\ll E_c$). In this limit, the 
backscatterer is weak. 
Charge transfer is then best understood by singling out two contributions:
First, transfer of  electrons that are not backscattered. They are responsible for   $I_0$.
Second, the contribution  due to backscattered electrons, generating $I_B$ which
converges to a limiting value $I_B^{\infty}=(e/4) T_B$ \cite{Kan92}. 
In the total current $I={\cal C}_1$, $I_B$ is, however, 
hidden under the contribution  $I_0=G_0 V/2 $ of unscattered electrons that increases 
linearly with $V$. Similarly, the large voltage limit 
${\cal C}^{\infty}_2 =  G_0 ( \pi T_B/4+T) $ of the second cumulant, 
Eq.\ (\ref{c2exakt}), has these two contributions: the  fluctuations due to the 
backscatterer as well as the Johnson-Nyquist noise 
$G_0 T$ of electrons that are not backscattered. Non-Gaussian current fluctuations, 
in contrast, allow to specifically probe the 
backscatterer and its large voltage behavior. This is because unscattered electrons 
produce purely Gaussian fluctuations. Accordingly we find that all higher order cumulants 
saturate at large 
voltages and, remarkably, their limiting values are temperature independent
(see Fig.\ \ref{fig_largeV} for ${\cal C}_3$). The 
generating function in this limit takes the form
\beq \label{wbs}
\ln {\cal Z}^{\infty}(\xi) = \tau\left[\frac{ G_0}{2} 
(-i  V \xi -  T \xi^2) + \frac{ T_B}{2} \left(e^{ie\xi/2}-1\right)\right].
\eeq
The temperature independence of higher order cumulants in the large voltage limit is in stark contrast to the behavior of a non-interacting conductor \cite{Lev93}. It is thus a clear signature of interactions in the QD considered here.  It can be understood by noting that a large voltage 
bias shifts the energies at which electron occupation numbers are thermally smeared 
far away from  the equilibrium Fermi level. Electrons with these high energies are, 
however, effectively not backscattered and therefore produce purely Gaussian noise. 
Thermal fluctuations do thus not contribute to higher order cumulants. 
With present day experimental techniques \cite{Reu03} this anomalous behavior can be 
observed in ${\cal C}_3$, the third cumulant of current fluctuations, Eq.\ (\ref{c3exakt}). 
It has the temperature independent large voltage limit 
${\cal C}^{\infty}_3=  -(e\pi /8)G_0 T_B$, as shown in Fig.\ \ref{fig_largeV}.
In the opposite limit of small voltages, we find at low temperatures  $T\ll eV \ll T_B$ (moderately low such that the inelastic processes that justify our model are still operative) the statistics 
\beq \label{sbs}
\ln {\cal Z}(\xi) = \tau \frac{ T_B}{48\pi}   \left(e^{-ie\xi}-1\right) \left
(\frac{eV}{T_B}\right)^3 +  {\cal O} \left(\frac{eV}{T_B}\right)^6.
\eeq
Due to the scaling of the impurity strength with energy, the high and the low voltage 
limits, Eqs.\ (\ref{wbs}) and (\ref{sbs}), correspond to the weak and the 
strong backscattering limit, respectively. They are Poissonian, in accordance with Refs.\ 
\cite{Lev01,Sal01}. At low voltages, as manifest in Eq.\ (\ref{sbs}), charge is 
transferred in units of the elementary charge. The current scales as $I\propto V^3$, 
as expected from  perturbative calculations \cite{Kan92}. 
The high voltage statistics, Eq.\ (\ref{wbs}),  suggests, that charge is transmitted in 
packets carrying {\it half} the elementary charge. This is a direct consequence of 
the charging interaction of the QD  that after 
every backscattering of an electron induces 
a positive electric potential on the QD.  It attracts electrons from the leads to 
compensate for the electron that is missing on the QD. In response, an electron  
flows onto the QD from either lead $1$ or 
from lead $2$. This  either cancels or completes the transfer of an electron through the 
QD that was initiated by the backscattering event. Both processes are equally likely 
in the weak backscattering limit of an almost open contact QPC $2$. Therefore only 
every other backscattering event transfers charge through the QD. Equivalently
one can say that every such event transfers only $e/2$, as indicated by 
Eq.\ (\ref{wbs}).

In the shot noise limit of zero temperature, the third cumulant for non-interacting 
electrons vanishes at transmission $\Gamma=1/2$. 
This is because in this case every electron is transmitted or reflected with the same 
probability $1/2$ independently of all other electrons. The distribution of transferred 
charge is consequently symmetric around its mean and its skewness ${\cal C}_3$ vanishes. 
This intuition 
remains correct for weakly interacting conductors \cite{Kin03}. One expects it to be 
invalidated, however, by strong interactions that correlate the transfers of different 
electrons  with each other. The inset of Fig.\ \ref{fig_largeV} shows that this is, 
indeed, the case for the QD we consider. We define an effective 
single-electron transmission 
$\Gamma_{\rm eff}= I/(G_0 V -I)$ for QPC $2$ in series with QPC $1$ by Ohm's law. 
${\cal C}_3$  vanishes then for $\Gamma_{\rm eff} \approx 0.26$, in clear contrast with a 
non-interacting structure.
 

In  conclusion, we have studied  current fluctuations in a strongly interacting 
quantum conductor at arbitrary transmission. While it had been found in Refs.\ \cite{Lev01,Kin03} that interactions do not qualitatively change the statistics of current fluctuations in perturbative situations, our nonperturbative solution does display 
features that are qualitatively different from those 
of   non-interacting structures. This makes the measurement of current correlations a promising tool to probe interactions in the QD we considered and most probably in many other strongly interacting conductors.  More specifically, we find that the fluctuations of the 
current  saturate at high voltages, while the current itself 
increases linearly with the applied voltage. Moreover, all cumulants higher than 
the second one are temperature independent in the high voltage limit.  
We have discussed in detail how  the voltage and temperature dependence of the third 
cumulant display these and other qualitative interaction effects. Experimental techniques 
for its measurement are available \cite{Reu03}. Our predictions are thus experimentally 
testable.   
 
We  thank P.\ W.\ Brouwer, H.\ Grabert, L.\ S.\ Levitov, K.\ A.\ Matveev, I.~Safi, 
and H.~Saleur for discussions. Financial support has been provided by the 
Cambridge-MIT Institute Ltd. and the EU network SPINTRONICS.


\begin{thebibliography}{99}
\bibitem{footnote}
Eq.\ (\ref{Zb}) can also be obtained 
from the Hamiltonian of a 
one-dimensional conductor of spinless electrons with backscattering and a 
series resistor of resistance $1/G_0$ along the lines of Ref.\ \cite{Saf03}. 
Our results apply, therefore, as well to the problem of the environmental 
Coulomb blockade with series resistance $1/G_0$.
\end{thebibliography}
\end{document}